\documentclass[journal]{IEEEtran}

\usepackage[]{graphicx}					% Supports jpg/png figure formats
\usepackage{amsmath,bm}					% Fancy math tools
\usepackage{esint}						% Provides extended sets of integral notation

\newcommand{\vecB}{{\bf B}}			% Magnetic field
\newcommand{\vecH}{{\bf H}}			% Auxiliary magnetic field
\newcommand{\vecJ}{{\bf J}}			% Electric current density
\newcommand{\vecM}{{\bf M}}			% Magnetization vector
\newcommand{\vecF}{{\bf F}}			% Force vector
\newcommand{\vecR}{{\bf r}}			% Position vector
\newcommand{\xHat}{\hat{\bf x}}		% Unit vector X
\newcommand{\yHat}{\hat{\bf y}}		% Unit vector Y
\newcommand{\zHat}{\hat{\bf z}}		% Unit vector Z
\newcommand{\rHat}{\hat{\bf r}}		% Unit vector R
\newcommand{\phiHat}{\hat{\bm \phi}}  % Unit vector phi
\begin{document}

% Paper title goes here.
\title{An Analytic Model for Eddy Current Separation}


\author{James~R.~Nagel \\ Department of Metallurgical Engineering \\ University of Utah, Salt Lake City, Utah % <-this % stops a space
%\thanks{James R. Nagel (james.nagel@utah.edu) is a research associate with the Department of Metallurgical Engineering at the University of Utah in Salt Lake City, Utah. His PhD in electrical engineering with an emphasis on numerical methods and applied electromagnetics.}% <-this % stops a space
}


%\markboth{IEEE TRANSACTIONS ON MAGNETICS}%

\maketitle

%==================================================================================================
% Fill abstract here. As a general rule, do not put math, special symbols or citations in the 
% abstract or keywords.
%==================================================================================================
\begin{abstract}
	Eddy current separation (ECS) is a process used throughout the scrap recycling industry for separating nonferrous metals from nonmetallic fluff. To date, however, the physical theory of ECS has generally been limited to empirical approximations and numerical simulations. We therefore introduce a simplified, two-dimensional model for ECS based on a cylindrical array of permanent, alternating magnets. The result is a Fourier-series expansion that describes the total magnetic field profile over all space. If the magnets are then rotated with constant angular velocity, the magnetic fields vary as a discrete series of sinusoidal harmonics, thereby inducing electrical eddy currents in nearby conductive particles. Force and torque calculations can then be used to predict the corresponding kinematic trajectories.
\end{abstract}

\begin{IEEEkeywords}
eddy currents, quasistatics, recycling
\end{IEEEkeywords}

% \fi
\IEEEpeerreviewmaketitle

%=========================================================================================
\section{Introduction}
%=========================================================================================

\IEEEPARstart{E}{ddy} current separation (ECS) is an integral process used throughout the recycling industry for recovering nonferrous metals from solid waste (e.g., Cu, Al, Zn), and also for separating nonferrous metals from each other~\cite{Schloemann75a, Schloemann75b, Schloemann82, Ruan14}. The technology operates on the principle that a time-varying magnetic field tends to induce electrical currents throughout the volume of a conductive particle. These so-called eddy currents (or Foucault currents) react to the applied magnetic field by exhibiting a distinct force of deflection that alters the kinematic trajectory of the metallic particle. If the deflection force is strong enough, it can even be used to rapidly separate highly-conductive metals (e.g. aluminum, copper, brass, etc) from other nonmetallic fluff (e.g. plastic, rubber, and glass). 

Despite the practical uses for eddy current technology, the basic theory of ECS is notoriously complex. Not only must one first be able to derive the magnetic field intensity $\vecB$ of some given magnetic configuration, but also the eddy current density $\vecJ$ excited by that field throughout some conductive particle geometry. Both of these tasks are unique mathematical challenges unto themselves, thus making it difficult to formulate a complete analytic model for the process. To date, the most popular solution has been to formulate the magnetic field profile in terms of a Fourier series of cylindrical harmonics and then approximate the coefficients through empirical measurements. The process seems to have originated with the work of Rem, et al~\cite{Rem97,Rem98}, and has since been applied by numerous other publications as well~\cite{Lungu02,Zhang99b,Maraspin04}. Alternatively, one may simply calculate the magnetic field profile using finite-element analysis and then apply a similar approximation numerically~\cite{Lungu01,Ruan11}.

%As a result, most prior attempts at modeling have tended to rely on a combination of numerical simulations and empirical approximations~\cite{Rem97, Rem98, Zhang99a, Zhang99b, Lungu02, Maraspin04, Ruan11}. 

What is needed is a closed-form mathematical model that completely solves for the magnetic field profile around an ECS without any reliance on computationally-intense simulations or measurement-based approximations. Not only will this greatly reduce the computational time required to model the trajectories of various metal particles, but it also provides mathematical insights into the physical nature of the process itself. Only by quantifying the various factors that contribute to particle separation can we efficiently explore new ways to optimize the technology for industrial applications.  

\begin{figure}
	\center
	\includegraphics[width = 2.5in]{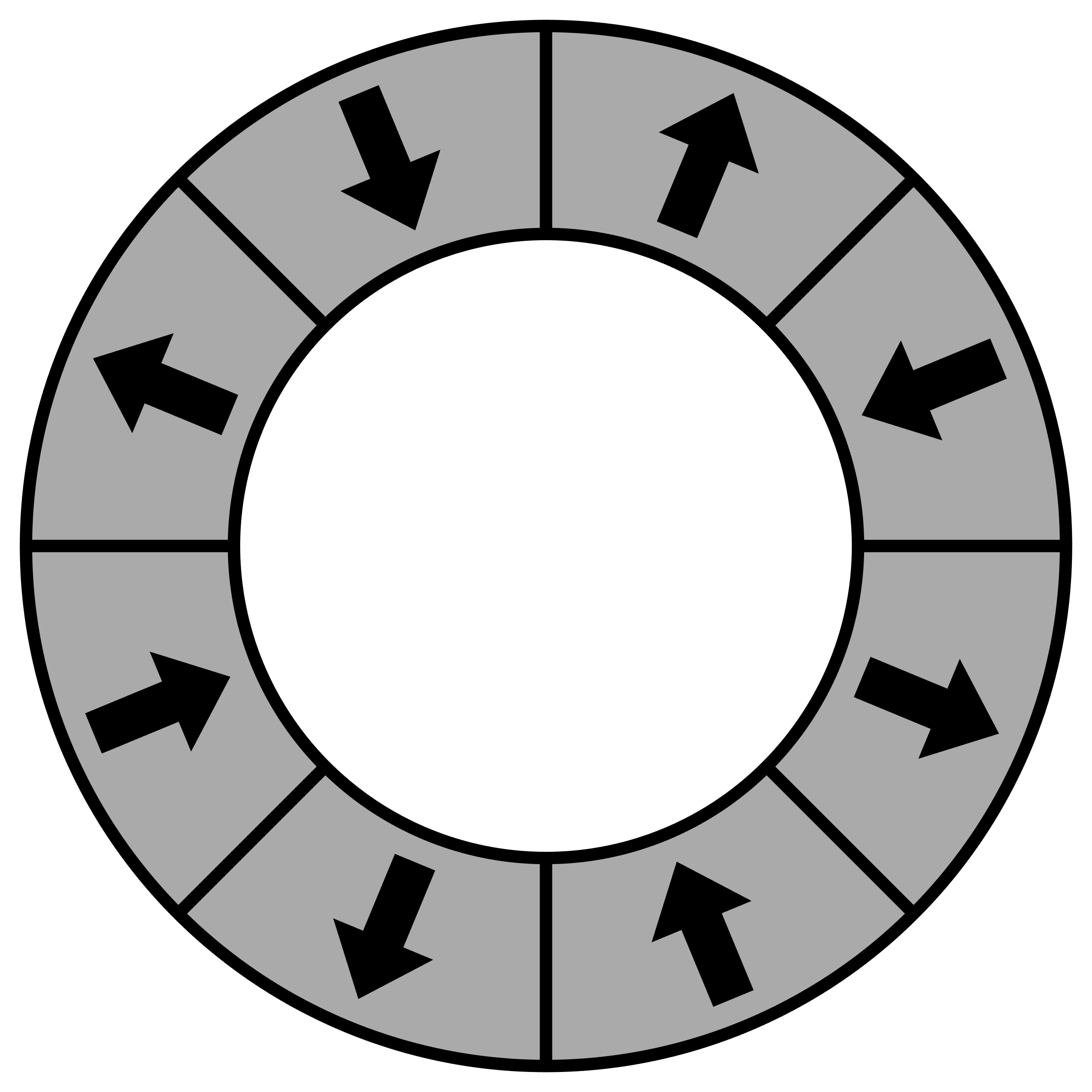}
	\caption{An eddy current separator represented as a cylindrical array of $K$ magnetized bars with alternating magnetization vector $\vecM$. The inner and outer radii are given as $R_a$ and $R_b$, respectively. \label{fig:Geometry} }
\end{figure}

Figure~\ref{fig:Geometry} shows an idealized depiction a typical eddy current separator that will be considered in this work. An array of magnetized bars are wrapped around in a closed, hollow cylinder with inner radius $R_a$ and outer radius $R_b$. The array may be treated as having infinite extent along $z$ so that $\partial / \partial z = 0$. The total number of bars is defined by an even number $K$, with $K = 16$ and $K = 32$ being common values for industrial separators. Each segment along the ECS is also assumed to have some permanent magnetization vector $\vecM$ pointing radially outward with alternating signs. 

Our first goal is to solve for the magnetic field profile $\vecB(x,y)$ throughout all space. We will then introduce a rotational velocity ${\bm \omega} = \omega \zHat$ and calculate the time-domain magnetic field profile around the bars. After that, we shall introduce a metal particle to the field and calculate the corresponding forces acting on it as it travels over the top of the ECS. This will allow us to finally track the trajectories of various metal particles as they are hurled away by the spinning magnets. 

%======================================================================
\section{Theoretical Background} \label{sec:Background}
%======================================================================

We begin with Maxwell's equations for linear, isotropic, nonmagnetic media in their static (zero frequency) formulation such that $d/dt = 0$. In particular, the static formulation of Ampere's law states that
\begin{equation} \label{eq:AmpereLaw}
	\nabla \times \vecB = \mu_0 \vecJ ~,
\end{equation} 

\noindent where $\vecB$ is the magnetic field intensity, $\vecJ$ is the electric current density, and $\mu_0$ is the permeability of free space. For the special case of a permanently magnetized material, the total current density $\vecJ$ represents the circulating flow of charge within the very atoms of the array. The phenomenon is inherently quantum mechanical in nature and arises primarily from the unbalanced spin of electrons in certain materials~\cite{Coey10, OHandley00}. Since no net charges can ever build up from this flow of current, we may conclude that $\vecJ$ has zero divergence and is thus solenoidal. We therefore express $\vecJ$ as curl of another vector field $\vecM$, called the magnetization field, such that 
\begin{equation}
	\vecJ = \nabla \times \vecM ~.
\end{equation}

%is expressed as a linear combination of two distinct sources, given as
%\begin{equation}
%	\vecJ = \vecJ_c + \vecJ_m ~.
%\end{equation}
%
%\noindent The $\vecJ_c$ term is called the conduction current density and represents the flow of charge in a metal due to an applied electric field $\vecE$. The behavior of this field is governed by the point form of Ohm's law, which states that
%\begin{equation}
%	\vecJ_c = \sigma \vecE ~,
%\end{equation}
%
%\noindent where $\sigma$ is the electrical conductivity of a given material. For a static system with no free charges, however, we can assume that $\vecE = \vecJ_c = 0$ throughout the magnetic bars.
%
%In contrast, the $\vecJ_m$ term is called the magnetization current and represents the circulating flow of charge within the very atoms of the magnetic bars. The origin of $\vecJ_m$ is inherently quantum mechanical in nature and arises primarily from the unbalanced spin of electrons in certain materials~\cite{Coey10, OHandley00}. Since no net charges can ever build up from this flow of current, we may conclude that $\vecJ_m$ has zero divergence, or $\nabla \cdot \vecJ_m = 0$. Any field which satisfies such a condition is said to be solenoidal and has the special property of being expressible as the curl of another vector field $\vecM$, called the magnetization field, such that 
%\begin{equation}
%	\vecJ_m = \nabla \times \vecM ~.
%\end{equation}

\noindent The physical interpretation of $\vecM$ is that of a net magnetization per unit volume. As such, the magnitude and direction of $\vecM$ is determined by the applied magnetic field that originally charged the magnet during fabrication.

Substitution into Ampere's law next reveals
\begin{equation}
	\nabla \times \left ( \vecB/\mu_0 - \vecM \right ) = 0 ~.
\end{equation}

\noindent We now introduce the auxiliary magnetic field $\vecH$ which is classically defined as\footnotemark
\begin{equation} \label{eq:Hfield}
	\vecH = \vecB/\mu_0 - \vecM ~. 
\end{equation}

\footnotetext{Note that we are using the modern naming conventions for the magnetic field $\vecB$ and the auxiliary magnetic field $\vecH$. See \cite{Arthur08} for more details.}

\noindent Since $\vecH$ is an irrotational vector field, we may infer the existence of a scalar field $\Phi$, called the magnetic scalar potential, such that $\vecH = - \nabla \Phi$. Substitution back into~\eqref{eq:Hfield} thus produces
\begin{equation}
	\nabla \Phi = \vecM -\vecB/\mu_0 ~.
\end{equation}

\noindent If we now take the divergence of both sides, then Gauss' law requires $\nabla \cdot \vecB = 0$ and we are left with
\begin{equation} \label{eq:Poisson}
	\nabla^2 \Phi = \nabla \cdot \vecM ~.
\end{equation}

\noindent We immediately recognize \eqref{eq:Poisson} as the well-known Poisson equation with $\nabla \cdot \vecM$ serving as the forcing function. As such, an analogous problem in electrostatics would be the electric scalar potential $V$ (i.e., voltage) arising from a volumetric charge density $\rho$. 

Since our present problem is cylindrical in symmetry, we may recall that the Laplacian in polar coordinates is given as
\begin{equation}
	\nabla^2 \Phi = \frac{1}{r} \frac{ \partial }{ \partial r} \left ( r \frac{ \partial \Phi }{ \partial r}  \right ) + \frac{1}{r^2} \frac{ \partial \Phi}{ \partial \phi^2} ~.
\end{equation}

\noindent where $(r,\phi)$ are the radial and angular coordinates. Using the method of separation of variables~\cite{Powers99}, the general solution to $\Phi(r,\phi)$ is known to satisfy
\begin{equation}
	\Phi = {\big [ } A r^{-\lambda} + B r^\lambda {\big ] } \big [ C \cos (\lambda \phi) + D \sin( \lambda \phi) \big ] ~,
\end{equation}

\noindent where $A$, $B$, $C$, and $D$ are arbitrary coefficients and $\lambda$ is the separation constant. Solutions to these constants depend on the boundary conditions over the domain of interest, which are addressed in the following sections.

%======================================================================
\section{Boundary Conditions} \label{sec:Boundaries}
%======================================================================

\begin{figure}
	\center
	\includegraphics[width = 2.0in]{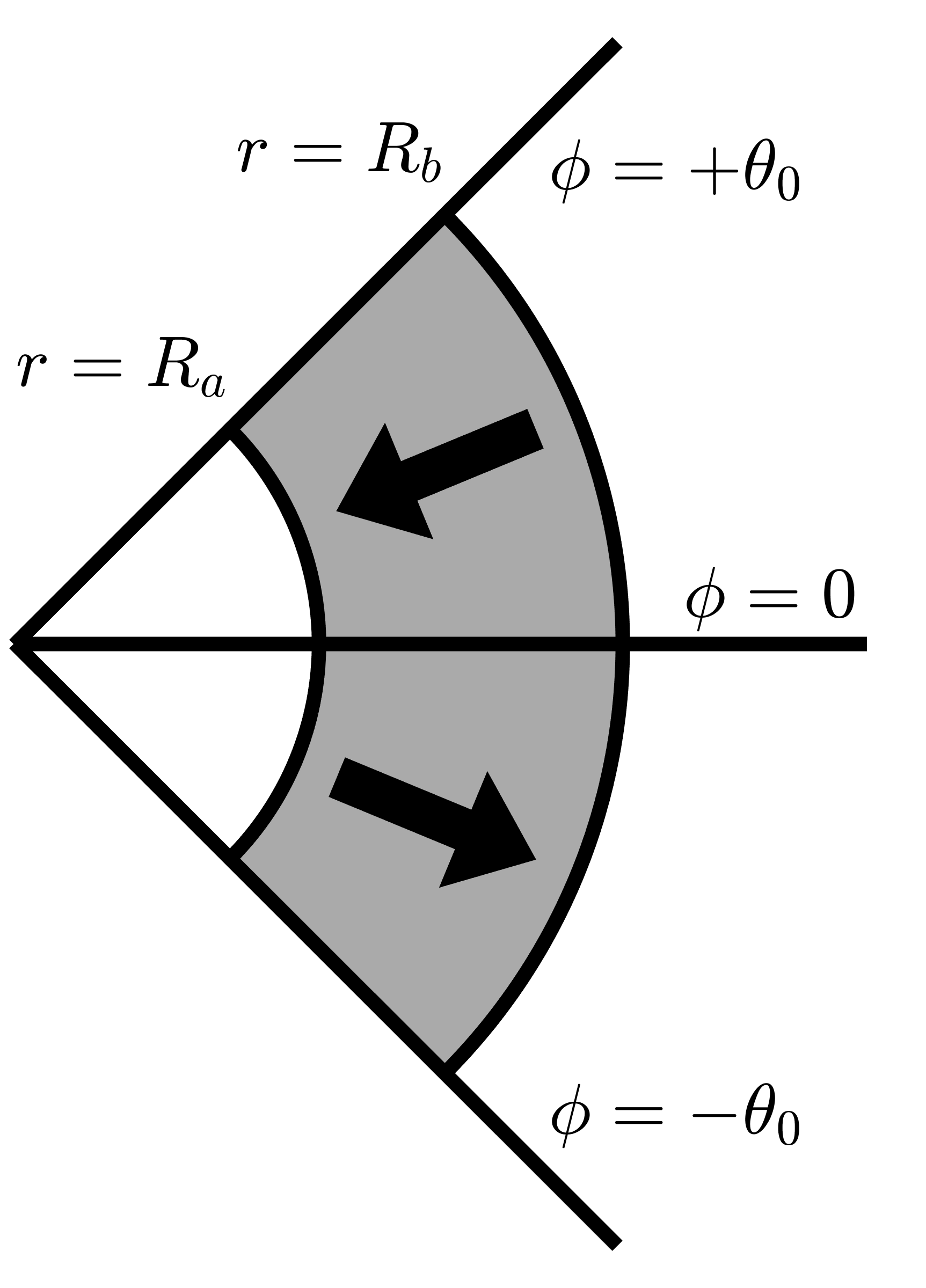}
	\caption{Unit cell of angular symmetry. The magnetic field profile repeats along arbitrary rotations of $\pm 2 \theta_0$. \label{fig:SinglePeriod} }
\end{figure}

Due to the angular symmetry of the system, we may limit our focus strictly to the region depicted in Figure~\ref{fig:SinglePeriod}. We can then impose periodicity along arbitrary rotations of $\pm 2 \theta_0$, where 
\begin{equation}
	\theta_0 = 2 \pi / K ~.
\end{equation}

\noindent Next, we need to impose a magnetization field $\vecM$ along the magnetic bars. Expressing this condition in polar coordinates, the region defined by $R_a \leq r \leq R_b$ is written as
\begin{equation}
	\vecM (r, \phi) = \rHat \frac{M_a R_a}{r}  \begin{cases} +1 & (-\theta_0 \leq \phi < 0) \\
	-1 & (0 < \phi \leq +\theta_0) \end{cases} ,
\end{equation}

\noindent where $\rHat = \vecR/|\vecR|$ is the radial unit vector. Outside of $R_a \leq r \leq R_b$, we simply let $\vecM = 0$.

The reason for formulating $\vecM$ in this way is to ensure that $\nabla \cdot \vecM$ is zero everywhere except for the discontinuities along $r = R_a$ and $r = R_b$. In practice, however, this may not be perfectly accurate since ferromagnetic materials are typically charged by applying a uniform magnetic field. Fortunately, if $R_a$ and $R_b$ are relatively large, then this distinction is not significant and the approximation should be reasonably accurate.

Using electrostatic theory as an analogy, we can imagine a mathematically equivalent system that is comprised of alternating strips of positive and negative surface charge density along $r = R_a$ and $r = R_b$. We may therefore express $\Phi$ as a linear superposition between two distinct contributions, labeled $\Phi_a$ and $\Phi_b$, representing the surface charges along the inner and outer radii. Each of these functions must then be split into two more distinct contributions representing outer and inner regions with respect to each surface. For example, we let $\Phi_a = \Phi_a^+ + \Phi_a^-$ to denote the contributions along $r > R_a$ and $r < R_a$, respectively. Likewise, $\Phi_b$ has two contributions, $\Phi_b^+$ and $\Phi_b^-$, defined by the regions $r > R_b$ and $r < R_b$. The total scalar potential $\Phi$ is then expressed as a linear superposition of all four terms, written as
\begin{equation}
	\Phi = \Phi_a^+ + \Phi_a^- + \Phi_b^+ + \Phi_b^- ~.
\end{equation}

As we shall find out shortly, the derivation of each potential function follows a nearly identical procedure. Thus, any solution found for one of them effectively gives us the solution for all four. We shall therefore focus our attention specifically on $\Phi_a^+$ and let the other three contributions follow naturally from basic transformations. 

Beginning with our first boundary condition, we simply require that $\Phi_a^+$ remain bounded as $r \to \infty$. The immediate consequence is thus $B = 0$. For our second boundary condition, we may further impose odd symmetry about the line along $\phi = 0$. This likewise forces $C = 0$, leaving the the more compact expression
\begin{equation}
	\Phi_a^+ = A r^{-\lambda} \sin( \lambda \phi)  \quad (r > R_a) ~. 
\end{equation}

\noindent For our third boundary condition, we apply periodicity in the form of
\begin{equation}
	\Phi_a^+(r,+ \theta_0) = \Phi_a^+(r,- \theta_0)  ~. 
\end{equation}

\noindent The immediate implication is that $\sin( \lambda \theta_0) = 0$, which can only be satisfied if
\begin{equation}
	\lambda_n = \frac{ n \pi }{ \theta_0 } = \frac{nK}{2} \quad (n = 1,\,2,\,3,\,\dots ) ~.
\end{equation}

\noindent Since any linear combination of solutions is also a solution, the general expression for $\Phi_a^+$ can now be written as
\begin{equation} \label{eq:PhiSeries_A}
	\Phi_a^+(r,\phi) = \sum_{n = 1}^{\infty} A_n r^{-\lambda_n} \sin( \lambda_n \phi) \quad(r > R_a) ~.
\end{equation}

\noindent We immediately recognize~\eqref{eq:PhiSeries_A} as a Fourier series along $\phi$ with polynomial decay along $r$. Following a similar argument for $\Phi_a^-$, the only difference is that $\Phi_a^- \to 0$ as $r \to 0$. The general expression for $\Phi_a^-$ is thus
\begin{equation} \label{eq:PhiSeries_B}
	\Phi_a^-(r,\phi) = \sum_{n = 1}^{\infty} A_n r^{+\lambda_n} \sin( \lambda_n \phi) \quad(r < R_a) ~.
\end{equation}

\noindent If we further impose continuity on $\Phi_a$ along $r = R_a$, then we find that the Fourier coefficients $A_n$ must be the same for both functions. 

To solve for the $A_n$ coefficients, we need to apply one final boundary condition along $r = R_a$. Unfortunately, the source expression $\nabla \cdot \vecM$ is somewhat challenging due to the discontinuity along this surface. A derivation for this condition is provided in Appendix~I, with the end result yielding
\begin{equation}
	\left . \frac{ \partial \Phi_a^+ }{ \partial r } \right |_{r = R_a} = -\frac{M_a}{2} \begin{cases} +1 & (-\theta_0 < \phi < 0) \\ -1 & (0 < \phi < +\theta_0) \end{cases} .
\end{equation}

\noindent Following standard methods of Fourier theory, we may take the derivative of~\eqref{eq:PhiSeries_A} with respect to $r$, multiply by $\sin(\lambda_m x)$, and then integrate over $[-\theta_0,\theta_0]$ to find
\begin{align}
	-\int \limits_{-\theta_0}^{+\theta_0} \sum_{n = 1}^{\infty} & \lambda_n A_n R_a^{-(\lambda_n + 1)} \sin( \lambda_n \phi ) \sin( \lambda_m \phi )\, d\phi \nonumber \\
	= - \frac{M_a}{2} \int \limits_{-\theta_0}^{0} & \sin( \lambda_m \phi ) \, d\phi + \frac{M_a}{2} \int \limits_{0}^{\theta_0} \sin( \lambda_m \phi ) \, d\phi ~.
\end{align}

\noindent By orthogonality, all summation terms on the left-hand side evaluate to zero except for $n = m$. After simplifying and solving for $A_n$, we quickly find that
\begin{equation}
	A_n = -\frac{2 M_a R_a^{\lambda_n+1}}{\lambda_n^2 \theta_0} \quad (n = 1,\,3,\,5,\,\dots ) ~. 
\end{equation}

\noindent The complete solution for $\Phi_a^+$ is now found to satisfy
\begin{equation}
	\Phi_a^+(r,\phi) = \sum_{n = 1}^{\infty} a_n \left ( \frac{R_a}{r} \right )^{\lambda_n} \sin( \lambda_n \phi) \quad (r > R_a)  ~,
\end{equation}

\noindent where $a_n = 2 M_a R_a / \lambda_n^2 \theta_0$. Note that the ratio $R_a/r$ will always be less than or equal to unity, thus guaranteeing convergence over the infinite series.

To account for the region defined by $r < R_a$, we simply follow a similar derivation for $\Phi_a^-$. Letting $\Phi_a = \Phi_a^+ + \Phi_a^-$, we find a compact solution satisfying
\begin{equation}
	\Phi_a(r,\phi) = \sum_{n = 1}^{\infty} a_n \sin( \lambda_n \phi) \left [ G_a(r) \right ]^{\lambda_n} ,
\end{equation}

\noindent where the radial gate function $G_a(r)$ is a piecewise function defined as
\begin{equation} \label{eq:Gate}
	G_a(r) = \begin{cases} r/R_a  & ( r \leq R_a) \\ R_a/r & (r \geq R_a) \end{cases} .
\end{equation}

\noindent To account for the contribution from the surface at $r = R_b$, we again repeat the same basic derivation to find
\begin{equation}
	\Phi_b(r,\phi) = \sum_{n = 1}^{\infty} b_n \sin( \lambda_n \phi) \left [ G_b(r) \right ]^{\lambda_n} ,
\end{equation}

\noindent where $b_n = -2 M_b R_b / \lambda_n^2 \theta_0$ and $M_b = M_a R_a/R_b$. The gate function $G_b$ likewise follows \eqref{eq:Gate} but with $R_b$ used in place of $R_a$. To find the total magnetic scalar potential over all space, we simply let $\Phi = \Phi_a + \Phi_b$ and arrive at
\begin{equation}
	\Phi(r,\phi) = \sum_{n = 1}^{\infty} \sin( \lambda_n \phi) \left [ a_n G_a(r)^{\lambda_n} + b_n G_b(r)^{\lambda_n} \right ] ~.
\end{equation}

%=========================================================================================
 \section{Magnetic Field Profile} \label{sec:Bfield}
%=========================================================================================

With a solution for $\Phi$ now in hand, we can solve for the auxiliary magnetic field intensity by letting $\vecH = - \nabla \Phi$. Begin by recalling that the gradient operation in polar coordinates satisfies
\begin{equation}
	\nabla \Phi = \frac{ \partial \Phi }{ \partial r} \rHat + \frac{1}{r} \frac{ \partial \Phi }{ \partial \phi } \phiHat ~.
\end{equation}

\noindent Letting $\vecH = H_r \rHat + H_{\phi} \phiHat$, we first solve for $H_r$ to find
\begin{align} \label{eq:Hr}
	H_r(r,\phi) = - & \sum_{n = 1}^{\infty}  \left ( \frac{\lambda_n}{r} \right ) \sin( \lambda_n \phi) \nonumber \\ 
	& \left [ a_n u_a(r) G_a(r)^{\lambda_n} + b_n u_b(r) G_b(r)^{\lambda_n} \right ]  ~, 
\end{align}

\noindent where $u_a$ and $u_b$ are step functions defined by 
\begin{equation} 
	u_a(r) = \begin{cases} +1 & ( r < R_a) \\ -1 & (r > R_a) \end{cases} ,
\end{equation}

\noindent with a similar expression for $u_b$. Calculating $H_{\phi}$ likewise produces
\begin{align} \label{eq:Hphi}
	H_{\phi}(r,\phi) = - & \sum_{n = 1}^{\infty} \frac{\lambda_n}{r} \cos( \lambda_n \phi) \nonumber \\ 
	& \left [ a_n G_a(r)^{\lambda_n} + b_n G_b(r)^{\lambda_n} \right ]  . 
\end{align}

\noindent To solve for magnetic field $\vecB$, we now add the magnetization field $\vecM$ to the H-field in accordance with
\begin{equation}
	\vecB = \mu_0 ( \vecH + \vecM ) ~.
\end{equation}

It is worth noting that, for most practical applications, we are generally only interested in the region defined by $r \geq R_b$ where $\vecM = 0$. In this case, the magnetic field simplifies to just $\vecB = \mu_0 \vecH$. It is also frequently convenient to convert the solution to rectangular coordinates by making use of the identities
\begin{align}
	B_x (r,\phi) &= \frac{x B_r(r,\phi) - y B_{\phi}(r,\phi) }{r} ~, ~\text{and} \\
	B_y (r,\phi) &= \frac{y B_r(r,\phi) + x B_{\phi}(r,\phi) }{r} ~,
\end{align}

\noindent where $r = \sqrt{x^2 + y^2}$. 

\begin{figure}
	\center
	\includegraphics[width = 3.5in]{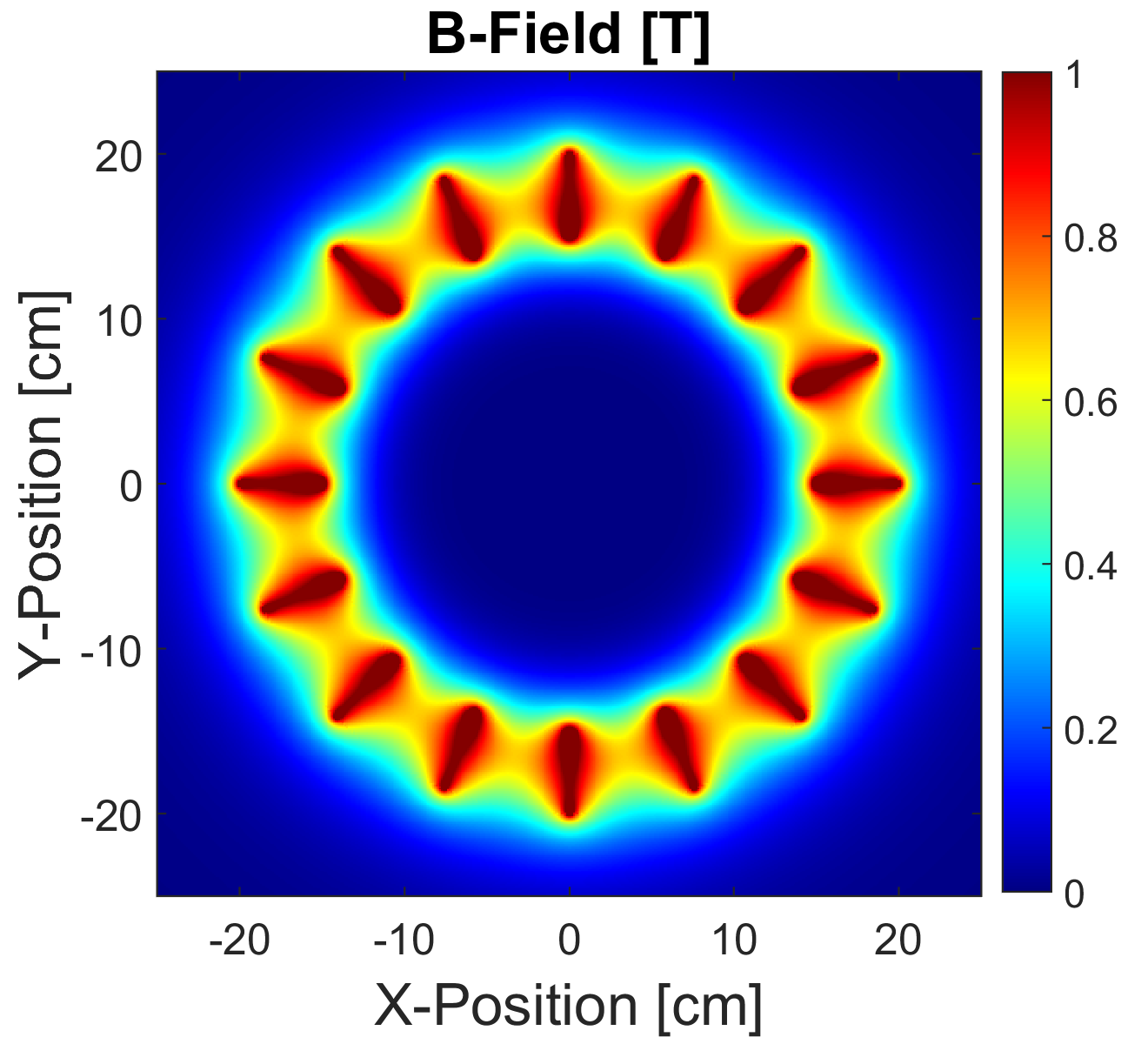} ~
	\includegraphics[width = 3.5in]{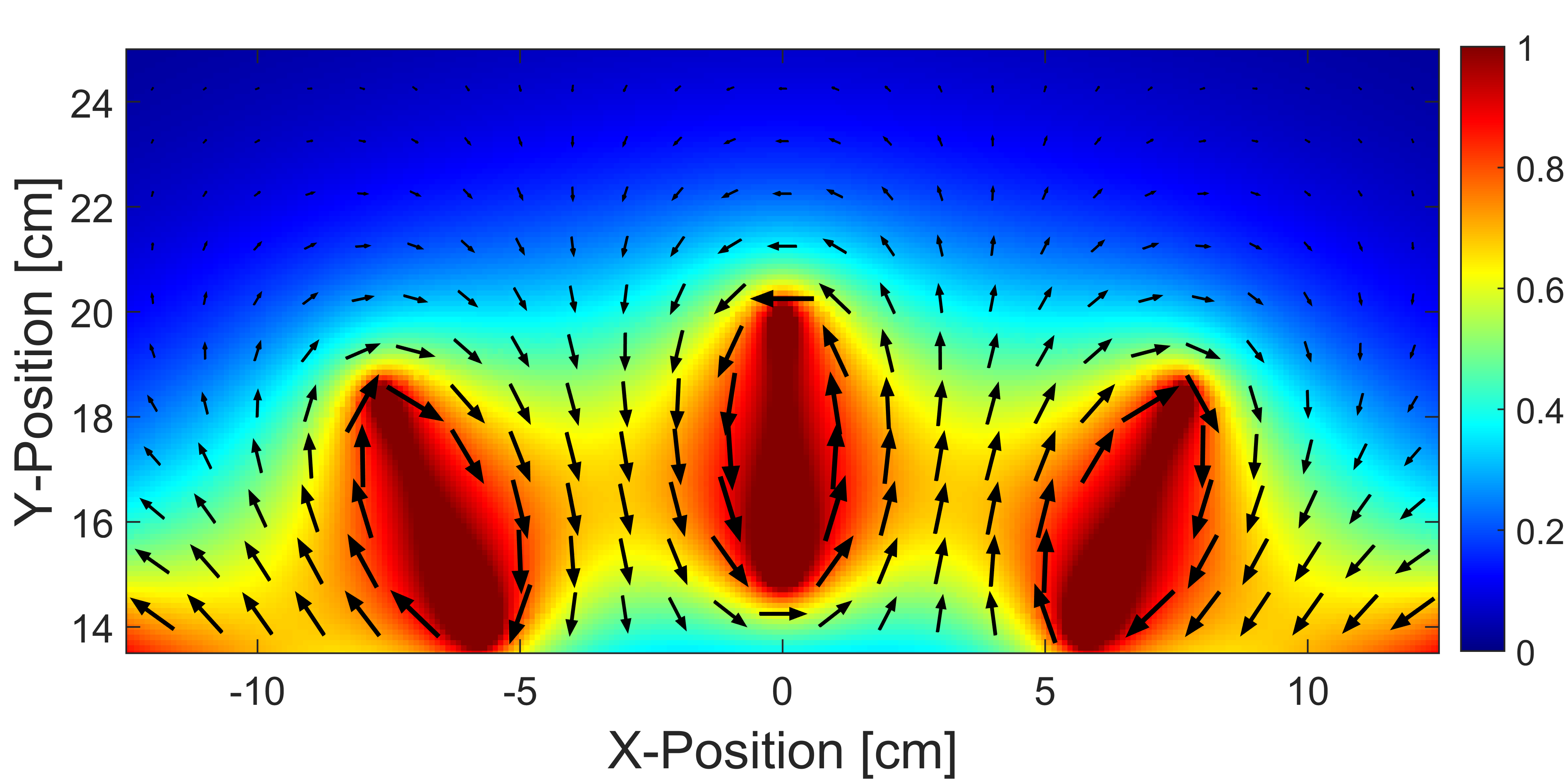}
	\caption{(Top) Magnetic field intensity for an eddy current separator with dimensions $R_a = 15~$cm and $R_b = 20~$cm with magnetization $M_a = 1.0~$MA/m. (Bottom) Close-up profile with vectors shown. \label{fig:BField} }
\end{figure}

Figure~\ref{fig:BField} shows an example calculation of the magnetic field intensity $\vecB$ for an eddy current separator with $K = 16$ poles, inner radius $R_a = 15~$cm, outer radius $R_b = 20~$cm, and internal magnetization $M_a = 1.0~$MA/m. Convergence of the Fourier series is surprisingly rapid, with only $N = 20$ terms producing highly accurate results. The only significant truncation errors seem to appear around the boundaries at $r = R_a$ and $r = R_b$ where the magnetization vector is discontinuous.

One interesting observation is that the contribution from the inner boundary at $R_a$ tends to negate the contribution at $R_b$. The implication is that if $R_a \to R_b$, then $\vecB \to 0$, which should be intuitive since an array with zero thickness produces no magnetic field. We also see that if the magnetic thickness $w = R_b - R_a$ is great enough, then the contribution from $R_a$ becomes negligible in the outer region $r \geq R_b$. This can be especially useful as an approximation for the magnetic field, which greatly simplifies into
\begin{align} \label{eq:Br}
	B_r(r,\phi) & \approx - \sum_{n = 1}^{\infty}  \alpha_n (r) \sin( \lambda_n \phi) ~, \text{~and} \\ \label{eq:Bp}
	B_{\phi}(r,\phi) & \approx + \sum_{n = 1}^{\infty}  \alpha_n (r) \cos( \lambda_n \phi)	~,
\end{align}

\noindent where the radial harmonic amplitudes satisfy
\begin{equation} \label{eq:HarmonicAmps}
	\alpha_n(r) =  \frac{2 \mu_0 M_b }{ n \pi } \left ( \frac{R_b}{r} \right )^{\lambda_n + 1} ~ (n = 1,\,3,\,5,\,\dots ) ~.
\end{equation}

\noindent Note that above formulation is perfectly consistent with the expressions provided by Rem, et al, in~\cite{Rem98}. The key difference, however, is that the Fourier coefficients are computed exactly from the underlying geometry and thus do not require any empirical measurements to satisfy as Rem and others have traditionally done.

One practical implication for eddy current separators is the desire to maximize $\vecB$ in the region beyond $r > R_b$ while simultaneously minimizing rotational inertia. Unfortunately, these are mutually incompatible goals. Maximization of $\vecB$ requires the array thickness $w$ to be very large while minimum rotational inertia requires $w$ to be very small. The ideal thickness, it seems, should be just enough to reasonably satisfy \eqref{eq:Br} and \eqref{eq:Bp}. Any extra thickness beyond this value is essentially dead weight, as it adds nothing to the overall field intensity outside of the array.

%=========================================================================================
 \section{Velocity Transformations} \label{sec:Velocity}
%=========================================================================================

To account for the relative motion between the cylindrical array and a piece of scrap metal overhead, we can imagine the magnetic field profile rotating with angular velocity ${\bm \omega} = \omega_0 \zHat$. This introduces a time dependence to the B-field written as 
\begin{equation} \label{eq:Spin}
	\vecB(r,\phi,t) = \vecB(r, \phi - \omega t) ~.
\end{equation}

\begin{figure}
	\center
	\includegraphics[width = 3.5in]{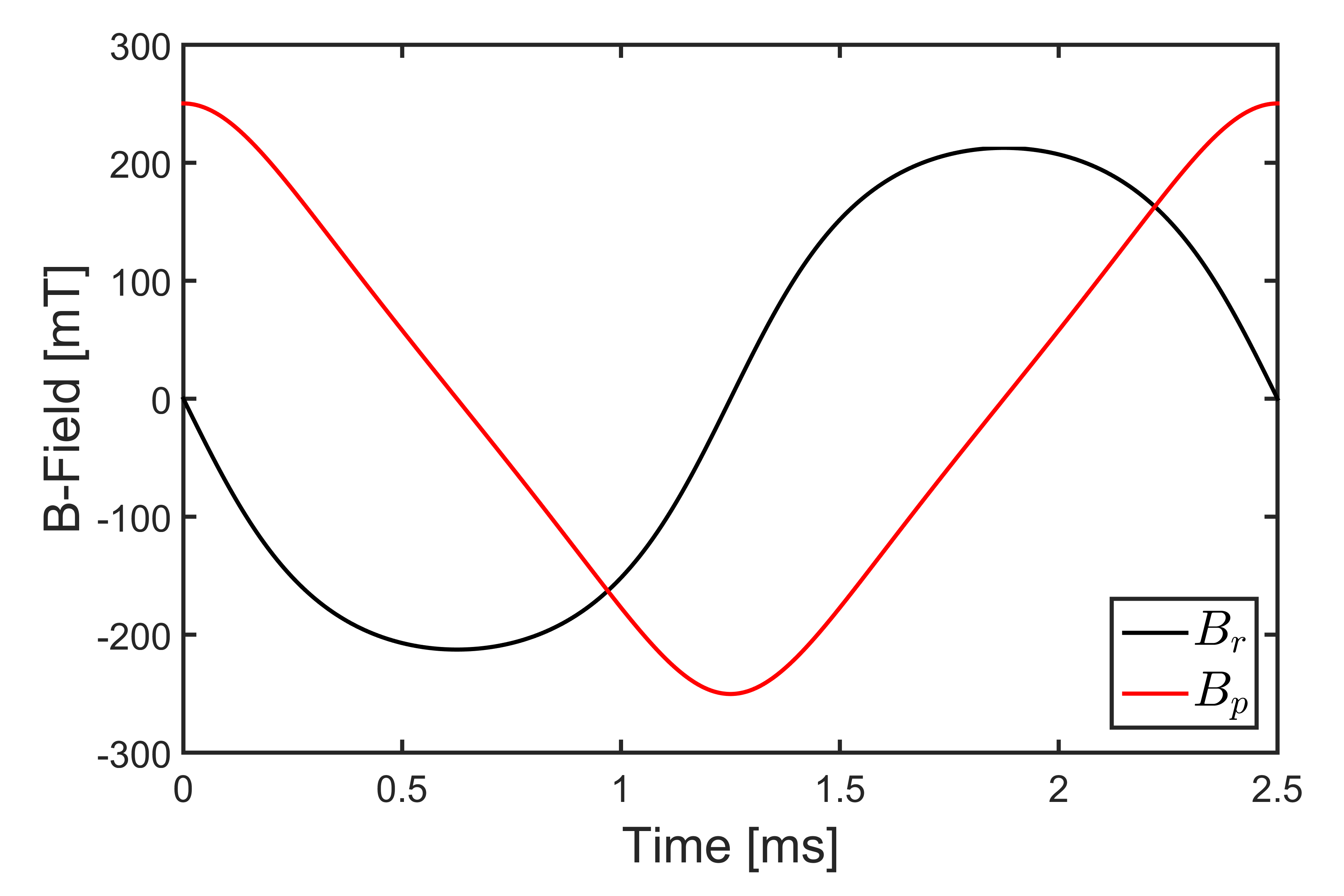} ~
	\caption{Time-domain signal for the magnetic field components if the magnetic array in Fig.~\ref{fig:BField} is rotated clockwise at 3000~rpm. The samples are taken at $x = 0$ and $y = 22~$cm.  \label{fig:TimeDomain} }
\end{figure}

\noindent Figure~\ref{fig:TimeDomain} shows a time-domain plot of the B-field components when the cylindrical array in Fig.~\ref{fig:BField} is rotated clockwise at 3000~rpm. The fields are sampled directly above the array at $(x,y) = (0,22)~$cm. From the perspective of a metal particle passing over the ECS, this plot represents the time-varying magnetic field profile that will induce eddy currents throughout its volume. 

The immediate implication of \eqref{eq:Spin} is that the B-fields can be expressed as a Fourer series of discrete temporal harmonics. Letting $j = \sqrt{-1}$, we first write each field component as
\begin{align} \label{eq:B_Fourier}
	B_r(r,\phi) &= \sum_{n = 1}^{\infty} \alpha_n(r) \, \text{Re} \left \{ j e^{j \lambda_n( \phi - \omega_0 t) } \right \} ~, ~\text{~and} \\
	B_{\phi}(r,\phi) &= \sum_{n = 1}^{\infty} \alpha_n(r) \, \text{Re} \left \{ e^{j \lambda_n( \phi - \omega_0 t) } \right \} ~, 
\end{align}

\noindent where Re$\{ x \}$ indicates the real part of $x$. We may now express each harmonic amplitude as a complex-valued phasor with the form
\begin{align} 
	\tilde{B}_r^n(r,\phi) &= j \alpha_n(r) e^{j \lambda_n \phi  } ~, ~\text{~and} \label{eq:B_Phasor_r} \\
	\tilde{B}_{\phi}^n(r,\phi) &= \alpha_n(r) e^{j \lambda_n \phi } ~.  \label{eq:B_Phasor_p}
\end{align}

\noindent Note that by convention, the time-dependence of each temporal harmonic is not expressly written, but merely implied. The angular frequency of each harmonic then satisfies $\omega_n = \lambda_n \omega_0$. Letting $\omega_n = 2 \pi f_n$, we may substitute for $\lambda_n$ and solve for $f_n$ to find 
\begin{equation} \label{eq:Harmonics}
	f_n = \frac{ n \omega_0}{ 2 \theta_0 } \quad (n = 1,\,3,\,5,\,\dots ) ~. 
\end{equation}

\begin{figure}
	\center
	\includegraphics[width = 3.5in]{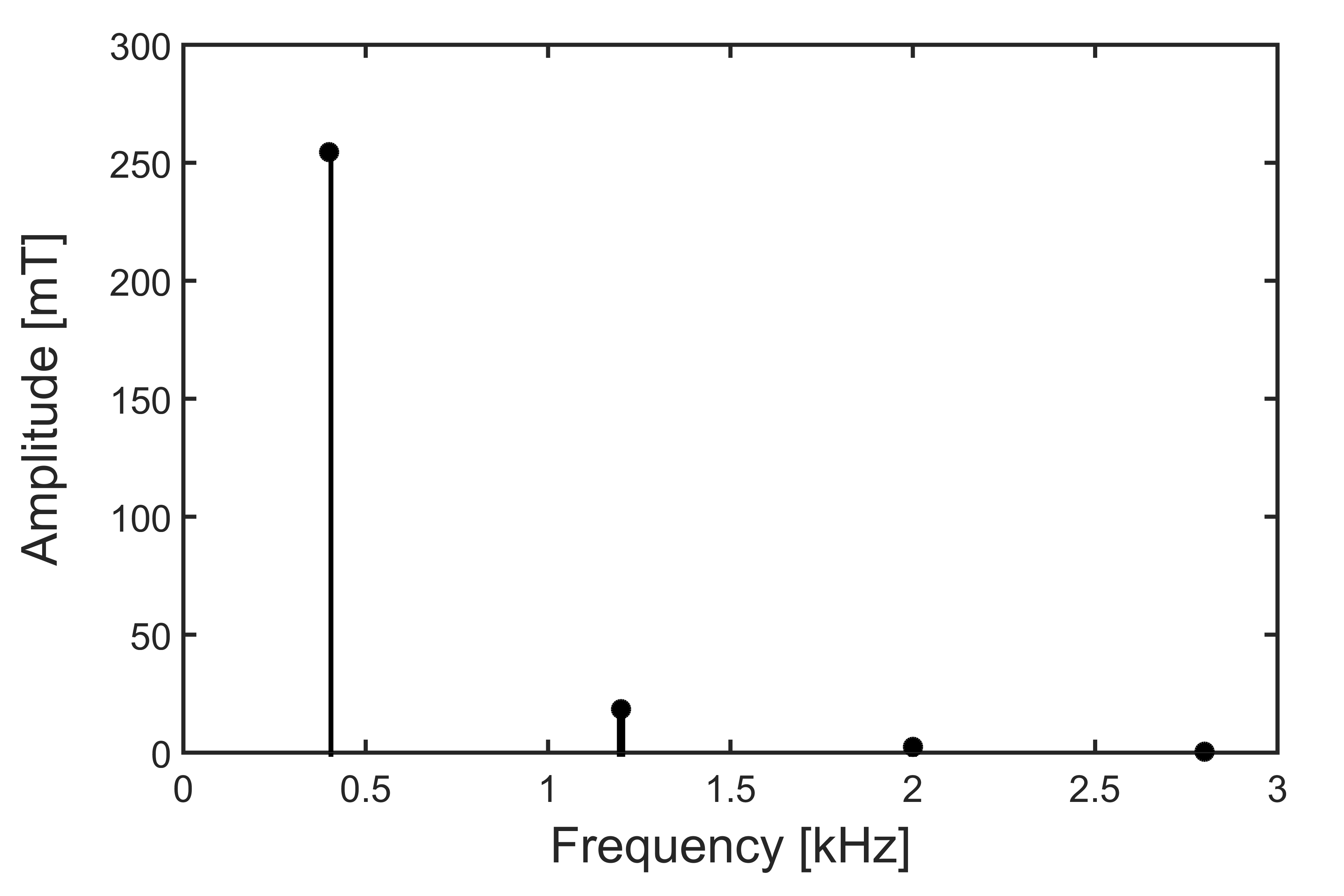} ~
	\caption{Approximate harmonic amplitudes for the time-domain signals depicted in Fig.~\ref{fig:TimeDomain}. \label{fig:Harmonics} }
\end{figure}

\noindent Figure~\ref{fig:Harmonics} shows the harmonic amplitudes of the time-domain signal when calculated using \eqref{eq:HarmonicAmps} and \eqref{eq:Harmonics}. The fundamental harmonic $(n = 1)$ clearly dominates the spectrum, though some moderate energy still persists in the higher frequencies. 

%=========================================================================================
 \section{Force Calculations} \label{sec:Force}
%=========================================================================================

To calculate the net force acting on a metal particle, we first imagine a small conductive sphere placed inside the time-varying magnetic field expressed by~\eqref{eq:Spin}. Due to the changing magnetic field, an electrical eddy current density $\tilde{\vecJ}$ will be induced throughout its volume, giving rise to a magnetic moment $\tilde{\bf m}$. Following the derivation in Appendix~II, a uniform, sinusoidal magnetic field $\tilde{\vecB}$ will give rise to a magnetic moment satisfying
\begin{equation} \label{eq:MagneticMoment}
	\tilde{\bf m} = \left ( \frac{ 6 \pi a^3 \tilde{\vecB} }{ \mu_0 } \right ) \left ( \frac{1}{k^2 a^2 } - \frac{\cot(ka)}{ka} - \frac{1}{3} \right ) ~,
\end{equation}

\noindent where $a$ is the spherical radius and $k = \sqrt{j \omega \sigma \mu_0}$. If we then introduce a small, linear gradient to the magnetic field $\tilde{\vecB}$, then the net, time-averaged force $\vecF_{avg}$ acting on the metal sphere can be shown to approximately satisfy~\cite{Jackson99}
\begin{equation}
	\vecF_{avg} = \frac{1}{2} \,  \text{Re} \left \{ \nabla ( \tilde{\bf m} \cdot \tilde{\vecB}^* ) \right \} ~.
\end{equation}

\noindent For the general case of non-spherical geometries, it has also been shown that this expression can be applied with good accuracy if an equivalent spherical radius is provided~\cite{Ray17}. We may therefore use it as a reasonable approximation for the behavior of many real-world scrap metal particles.

If the linear gradient on $\tilde{\vecB}$ is relatively minor, then the magnetic moment $\bf m$ may be treated as a constant value arising from the average magnetic field throughout its volume. This allows us to expand out the gradient in cylindrical coordinates such that
\begin{align}
	\nabla ( \tilde{\bf m} \cdot \tilde{\vecB}^*) = \left ( \tilde{m}_r \frac{\partial \tilde{B}_r^*}{ \partial r} + \tilde{m}_{\phi} \frac{\partial \tilde{B}_{\phi}^*}{ \partial r} \right ) \,  & \rHat \nonumber \\
	+ \frac{1}{r} \left ( \tilde{m}_r \frac{\partial \tilde{B}_r^*}{ \partial \phi} + \tilde{m}_{\phi} \frac{\partial \tilde{B}^*_{\phi}}{ \partial \phi} \right ) \, & \phiHat ~.
\end{align}

\noindent If we next make use of the identity
\begin{equation}
	\frac{ \partial}{\partial r} \alpha_n(r) =  - \left ( \frac{ \lambda_n +1 }{r} \right ) \alpha_n(r) ~,
\end{equation}

\noindent we can then calculate the derivatives of~\eqref{eq:B_Phasor_r} and~\eqref{eq:B_Phasor_p} to find
\begin{align}
	\frac{\partial \tilde{B}^n_r}{ \partial r} &= - \left ( \frac{ \lambda_n +1 }{r} \right ) \tilde{B}^n_r ~, \\
	\frac{\partial \tilde{B}^n_{\phi}}{ \partial r} &= - \left ( \frac{ \lambda_n +1 }{r} \right ) \tilde{B}^n_{\phi} ~, \\
	\frac{1}{r} \frac{\partial \tilde{B}^n_r}{ \partial \phi} &= \left ( \frac{j \lambda_n}{r} \right ) \tilde{B}^n_r ~, \\
	\frac{1}{r} \frac{\partial \tilde{B}^n_{\phi}}{ \partial \phi} &= \left ( \frac{j \lambda_n}{r} \right ) \tilde{B}^n_{\phi} ~.
\end{align}

\noindent Since the fundamental harmonic apparently dominates the magnetic field spectrum, it should be reasonable to calculate $\vecF_{avg}$ from this term alone. If greater accuracy is desired, however, then all one need do is add up the individual forces over many discrete harmonics. For even further accuracy still, one could also drop the approximations of \eqref{eq:Br} and~\eqref{eq:Bp}. This would be especially important for thinner magnetic arrays where $R_a$ is relatively close to $R_b$. 

%=========================================================================================
 \section{Kinematic Simulations} \label{sec:Sims}
%=========================================================================================

\begin{figure}
	\center
	\includegraphics[width = 3.5in]{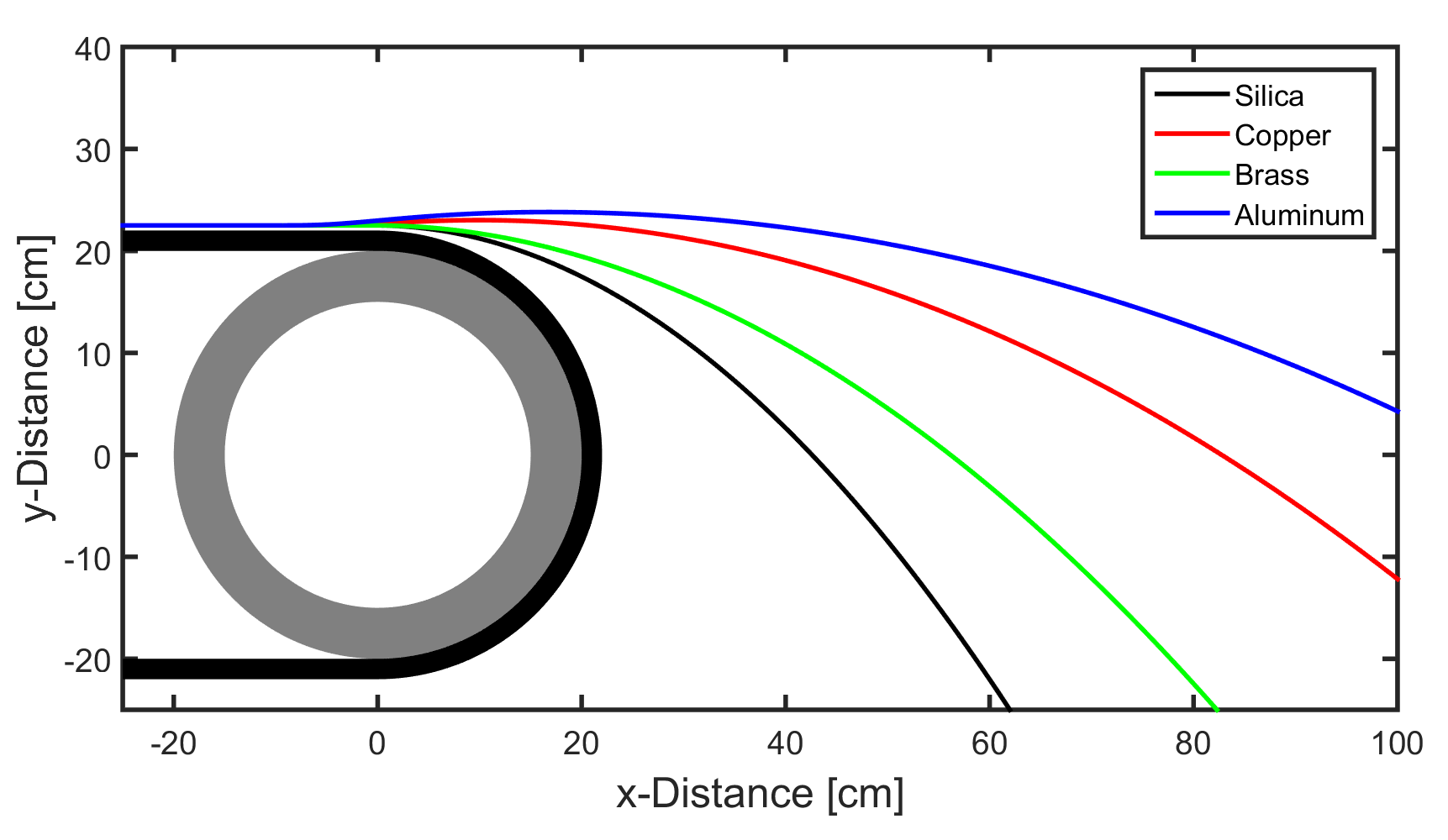}
	\caption{Kinematic trajectories for various metal spheres ($a = 0.5$~cm) passing over the ECS from Fig.~\ref{fig:BField}. The conveyor belt is moving with a constant horizontal velocity of 2.0~m/s while the array spins clockwise at 3000~rpm. \label{fig:Trajectories} }
\end{figure}

Once we are able to calculate the net force acting on a metal particle, it is finally possible to construct a full kinematic trajectory as it passes over the ECS. Depicted in Fig.~\ref{fig:Trajectories}, we may imagine a sampling of various metal spheres with radius $a = 0.5$~cm as they travel through the magnetic field along a moving conveyor belt. The space between the magnet and the belt was assumed to be 2.0~cm with the belt traveling at a constant horizontal velocity of 2.0~m/s. For simplicity, we may also neglect any frictional forces between the belt and the metal particles as well as air resistance. Thus, the only forces of significance are gravity, the normal force of contact with the belt, and the magnetic force due to the induced eddy currents.

To simulate the kinematic trajectories of each particle, we simply calculated the force, acceleration, velocity, and position over small time increments of 0.5~ms. After each increment, a new position can be derived from the updated parameters by applying Newton's laws of motion. For simplicity, we may neglect the contributions due to relative motion between the metal particles and the spinning array. At $\omega = 3000$~rpm, the rotational velocity at the edge of the magnetic array is nearly $63$~m/s, thus dwarfing the 2.0~m/s of horizontal velocity along the conveyor.

Tab.~\ref{tab:Materials} summarizes the four material compositions demonstrated by the model (silica, copper, brass, and aluminum). Since silica has essentially zero conductivity, its trajectory through the ECS represents the natural path taken by any free-falling body with no response to the magnetic field. The other metals have varying degrees of conductivity and density which all respond to the spinning magnetic array differently. Brass, being relatively dense with low conductivity, tends to throw very slightly in the ECS. Copper, however, has much greater conductivity and thus experiences a significantly greater deflection. Aluminum is likewise very conductive, but also much lower in density than either copper or brass. As such, we see that aluminum experiences the greatest throw distance of all. 
\begin{table}
	\begin{center}
	\caption{Electrical conductivity and mass density for various metals under consideration. \label{tab:Materials}}
	\begin{tabular}{l || c c} \hline
		Material & Conductivity~[MS/m] & Density~[g/cm$^3$] \\ \hline
		Silica & $\approx 0$ & 2.7  \\  
		Copper & 58.5 & 9.0  \\ 
		Brass & 15.9 & 8.5  \\ 
		Aluminum & 34.4 & 2.7  \\  \hline 
	\end{tabular}
	\end{center}
\end{table}

%=========================================================================================
\section{Conclusions} \label{sec:Conclusion}
%=========================================================================================

This paper provides a set of closed-form expressions for the magnetic field profile surrounding a cylindrical array of permanent magnets like that of a typical eddy current separator. The results are congruent with the assumptions of previous publications which expressed the field profile as a Fourier series of angular harmonics in cylindrical coordinates. Rather than empirically approximate the coefficients, however, it is now possible to calculate them analytically from the physical parameters of the model. 

%=========================================================================================
\section*{Appendix I} 
%=========================================================================================

\begin{figure}
	\center
	\includegraphics[width = 3.0in]{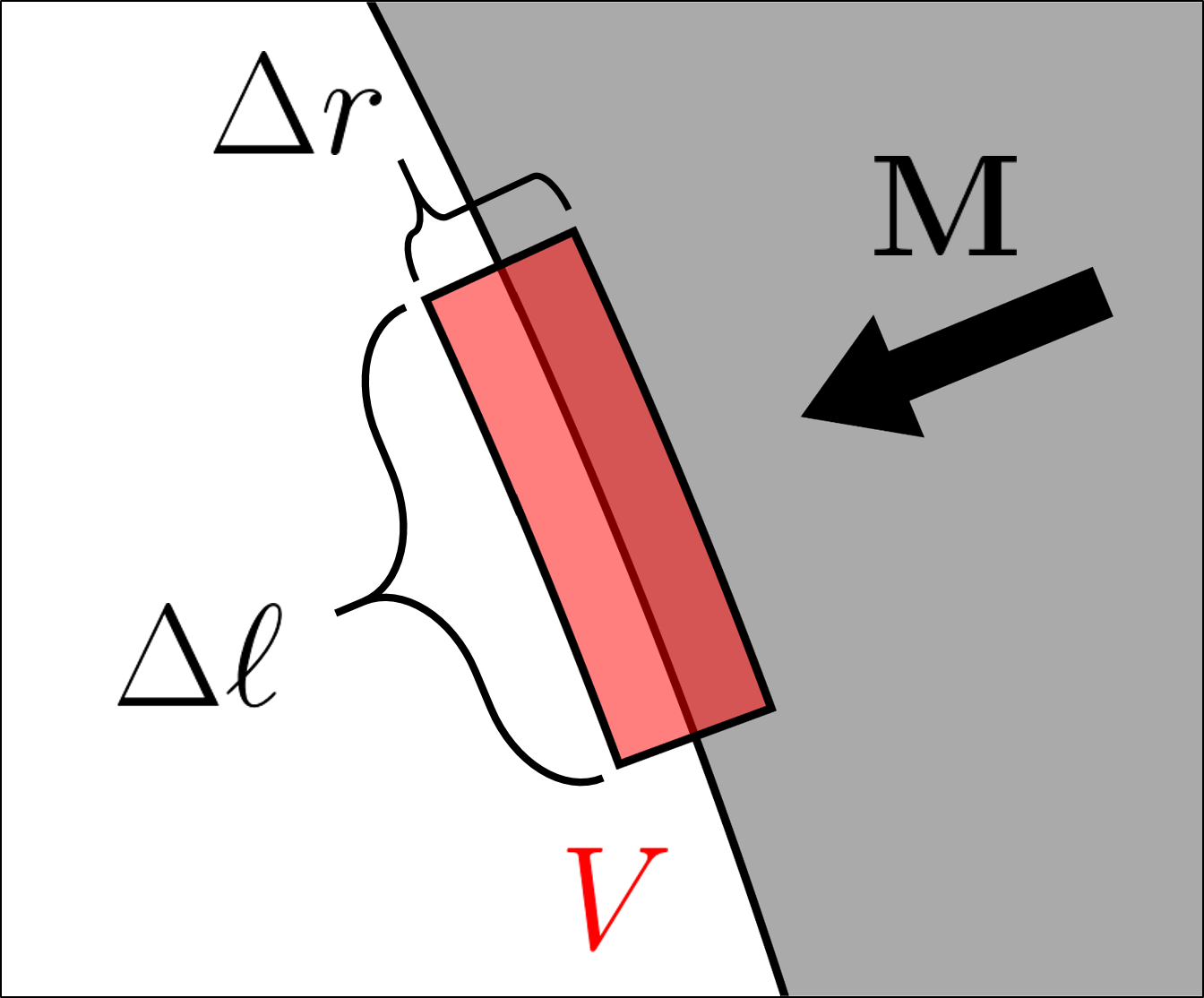}
	\caption{The volume of integration $V$ used for deriving the boundary condition on the potential function $\Phi_a^+$ at $r = R_a$. \label{fig:Pillbox} }
\end{figure}

The following section derives the boundary condition on $\Phi_a^+$ at $r = R_a$. We begin by imagining the tiny volume $V$ depicted in Fig.~\ref{fig:Pillbox} that straddles the boundary along $r = R_a$ at some angle far away from $\phi = 0$. If we calculate the volume integral of \eqref{eq:Poisson} over $V$, we find that 
\begin{equation}
	\iiint \limits_V \nabla^2 \Phi \, dV = \iiint \limits_V \nabla \cdot \vecM \, dV ~.
\end{equation}

\noindent By applying the divergence theorem, the volume integrals turn into surface integrals with the form
\begin{equation} \label{eq:Surfaces}
	\oiint \limits_S \nabla \Phi \cdot d {\bf S} = \oiint \limits_S \vecM \cdot d {\bf S} ~,
\end{equation}

\noindent where $S$ now denotes the surface enclosing $V$ and $d {\bf S}$ indicates the outward-pointing differential unit normal. Since $\vecM \cdot d {\bf S}$ is zero everywhere except for the outside face, the right-hand side immediately evaluates to 
\begin{equation}
	\oiint \limits_S \vecM \cdot d {\bf S} = -M_a \Delta \ell \begin{cases} +1 & (-\theta_0 < \phi < 0) \\ -1 & (0 < \phi < +\theta_0) \end{cases} ,
\end{equation}

\noindent where $\Delta \ell \approx R_a \Delta \phi$.

To calculate the left-hand side of \eqref{eq:Surfaces}, we must further include the nonzero contributions from the other three faces of the surface $S$. Since $\Delta \phi$ and $\Delta r$ are very small, we can approximate this result as 
\begin{align}
	\oiint \limits_S \nabla \Phi \cdot d {\bf S} \approx  \Delta \ell \frac{ \partial }{\partial r} & \left [ \Phi_a^+(R_a,\phi) - \Phi_a^-(R_a,\phi) \right ] \nonumber \\ + \frac{\Delta r}{R_a} \frac{ \partial }{\partial \phi} & \left [ \Phi_a^+(R_a,\phi) - \Phi_a^-(R_a,\phi) \right ] . 
\end{align}

\noindent By imposing continuity along $r = R_a$, we must conclude that $\Phi_a^+(R_a,\phi) = \Phi_a^-(R_a,\phi)$. If we further examine \eqref{eq:PhiSeries_A} and \eqref{eq:PhiSeries_B}, we find an even symmetry about the angular derivatives and an odd symmetry about the radial derivatives such that
\begin{align}
	\frac{ \partial }{\partial r} \Phi_a^+(R_a,\phi) & = - \frac{ \partial }{\partial r} \Phi_a^-(R_a,\phi)  ~, \\ 
	\frac{ \partial }{\partial \phi} \Phi_a^+(R_a,\phi) & = + \frac{ \partial }{\partial \phi} \Phi_a^-(R_a,\phi) ~. 
\end{align}

\noindent Consequently, the angular derivatives sum to zero while the radial derivatives add together, giving
\begin{equation}
	\oiint \limits_S \nabla \Phi \cdot d {\bf S} \approx 2 \Delta \ell \frac{ \partial }{\partial r} \Phi_a^+(R_a,\phi) ~. 
\end{equation}

\noindent If we now take the limit as $(\Delta r,\Delta \phi) \to 0$, the $\Delta \ell$ terms cancel and the approximations become exact. The final boundary condition thus satisfies
\begin{equation}
	\frac{ \partial }{ \partial r } \Phi_a^+(R_a,\phi) = -\frac{M_a}{2} \begin{cases} +1 & (-\theta_0 < \phi < 0) \\ -1 & (0 < \phi < +\theta_0) \end{cases} . 
\end{equation}

\noindent Following the same argument, it is straightforward to show that the boundary condition on $\Phi_a^-$ likewise satisfies
\begin{equation}
	\frac{ \partial }{ \partial r } \Phi_a^-(R_a,\phi) = +\frac{M_a}{2} \begin{cases} +1 & (-\theta_0 < \phi < 0) \\ -1 & (0 < \phi < +\theta_0) \end{cases} ,
\end{equation}

\noindent with similar expressions following for $\Phi_b^+$ and $\Phi_b^-$ along $r = R_b$. 

%=========================================================================================
\section*{Appendix II} 
%=========================================================================================

According to the derivation of~\cite{Nagel18}, a metal sphere with radius $a$ and conductivity $\sigma$ will exhibit an eddy current density $\tilde{\vecJ}$ when placed in a time-varying magnetic field $\tilde{\vecB} = \tilde{B}_0 \zHat$. Assuming an angular frequency $\omega$ of excitation, the current density evaluates to 
\begin{equation}
	\tilde{\vecJ} (r,\theta) = \phiHat \left ( \frac{ - 3 j \omega \sigma a \tilde{B}_0 }{ 2 k a j_1^{\prime} (ka) + 4 j_1 (ka) } \right ) j_1 (kr) \sin \theta ~,
\end{equation}

\noindent  where the function $j_n(x)$ is the spherical Bessel function of the first kind with order $n$ (not to be confused with the imaginary unit $j = \sqrt{-1}$). To conform with \eqref{eq:Spin}, however, we must adopt a phasor convention of $d/dt = -j\omega$ rather than $+j \omega$. This has the result of slightly modifying the wavenumber $k$ such that $k = \sqrt{j \omega \sigma \mu_0}$ rather than $k = \sqrt{-j \omega \sigma \mu_0}$. 

Our first goal is to simplify $\tilde{\vecJ}$. To accomplish that task, we require the following identities:
\begin{align}
	j_1^{\prime}(x) &= \frac{ j_1(x) }{ x} - j_2 (x) ~, \\
	j_2 (x) &= \left (\frac{3}{x} \right ) j_1(x) - j_0 (x) ~, \\
	j_0(x) &= \frac{ \sin(x) }{ x } ~, \\
	j_1(x) &= \frac{ \sin(x) }{ x ^2} - \frac{ \cos(x) }{ x } \label{eq:j_1}~.
\end{align}

\noindent This allows us to rewrite the leading coefficient on $\tilde{\vecJ}$ such that
\begin{equation}
	\tilde{\vecJ} (r,\theta) = \phiHat \left ( \frac{ - 3 j \omega \sigma a \tilde{B}_0 }{ 2 \sin (ka)  } \right ) j_1 (kr) \sin \theta
\end{equation}

Our next goal is to calculate the magnetic moment $\tilde{\bf m}$, defined as
\begin{align}
	\tilde{\bf m} &= \frac{1}{2} \int_V \vecR \times \vecJ \, dV \nonumber \\ 
	&= \int \limits_0^{2 \pi} \int \limits_0^{\pi} \int \limits_0^a ( r \rHat \times \phiHat ) \tilde{J}_{\phi} (r,\theta) r^2 \sin \theta \, dr d\theta d \phi ~.
\end{align}

\noindent The cross product $\rHat \times \phiHat$ is a function of position and needs to be treated with care. When expressed in rectangular unit vectors, we find that
\begin{equation}
	\rHat \times \phiHat = - \xHat \cos \theta \cos \phi - \yHat \cos \theta \sin \phi + \zHat \sin \theta ~. 
\end{equation}

\noindent The integrals over $\cos \phi$ and $\sin \phi$ both evaluate to zero, leaving only the $\zHat$ component. The integral over $\phi$ then evaluates to $2 \pi$, leaving 
\begin{equation}
	\tilde{\bf m} = \zHat \left ( \frac{ - 6 \pi j \omega \sigma a \tilde{B}_0 }{ 2 \sin (ka)  } \right ) \int \limits_0^{\pi} \int \limits_0^a j_1 (kr) r^3 \sin^3 \theta \, dr d\theta ~.
\end{equation}

\noindent The integral over $\sin^3 \theta$ is also straightforward to solve and evaluates to $4/3$. The remaining integral requires use of \eqref{eq:j_1} and evaluates to
\begin{equation}
	\int \limits_0^a j_1 (kr) r^3 \, dr = \frac{ (3 - k^2 a^2 ) \sin(ka) - 3ka \cos(ka)}{k^2} ~.
\end{equation}

\noindent After some simplification, the magnetic moment can then be shown to satisfy
\begin{equation} \label{eq:MagMoment}
	\tilde{\bf m} = \left ( \frac{ 6 \pi a^3 \tilde{\vecB} }{ \mu_0 } \right ) \left ( \frac{1}{k^2 a^2 } - \frac{\cot(ka)}{ka} - \frac{1}{3} \right ) ~,
\end{equation}

\noindent where $\tilde{\vecB} = \tilde{B}_0 \zHat$. More generally, however, we can orient $\tilde{\vecB}$ along any arbitrary direction and \eqref{eq:MagMoment} would still be valid. 

%=========================================================================================
 \section*{Acknowledgments} \label{sec:Thanks}
%=========================================================================================
 
This work was funded by the United States Advanced Research Project Agency-Energy (ARPA-E) METALS Program under cooperative agreement grant DE-AR0000411. The author would also like to thank Professor Raj Rajamani and Dawn Sweeney for their insightful edits and proof-reading.

%================================================================================================%  Bibliography
%==================================================================================================
%\bibliographystyle{IEEEtran}
\bibliographystyle{ieeetr}
\bibliography{ECSBib}

\begin{thebibliography}{10}

\bibitem{Schloemann75a}
E.~Schloemann, ``Separation of nonmagnetic metals from solid waste by permanent
  magnets. {I}. theory,'' {\em Journal of Applied Physics}, vol.~46, no.~11,
  pp.~5012--5021, 1975.

\bibitem{Schloemann75b}
E.~Schloemann, ``Separation of nonmagnetic metals from solid waste by permanent
  magnets. {II}. experiments on circular discs,'' {\em Journal of Applied
  Physics}, vol.~46, no.~11, pp.~5022--5029, 1975.

\bibitem{Schloemann82}
E.~Schloemann, ``Eddy-current techniques for segregating nonferrous metals from
  waste,'' {\em Conservation \& recycling}, vol.~5, no.~2--3, pp.~149--162,
  1982.

\bibitem{Ruan14}
J.~Ruan, Q.~Yiming, and Z.~Xu, ``Environmentally-friendly technology for
  recovering nonferrous metals from e-waste: Eddy current separation,'' {\em
  Resources, Conservation and Recycling}, vol.~87, pp.~109--116, June 2014.

\bibitem{Rem97}
P.~C. Rem, P.~A. Leest, and A.~J. van~den Akker, ``A model for eddy current
  separation,'' {\em International Journal of Mineral Processing}, vol.~49,
  pp.~193--200, 1997.

\bibitem{Rem98}
P.~C. Rem, E.~M. Buender, and A.~J. van~den Akker, ``Simulation of eddy current
  separators,'' {\em IEEE Transactions on Magnetics}, vol.~34, no.~4,
  pp.~2280--2286, 1998.

\bibitem{Lungu02}
M.~Lungu and P.~Rem, ``Separation of small nonferrous particles using an
  inclined drum eddy-current separator with permanent magnets,'' {\em {IEEE}
  Transactions on Magnetics}, vol.~38, no.~3, pp.~1534--1538, 2002.

\bibitem{Zhang99b}
S.~Zhang, P.~C. Rem, and E.~Forssberg, ``Particle trajectory simulation of
  two-drum eddy current separators,'' {\em Resources, Conservation and
  Recycling}, vol.~26, pp.~71--90, 1999.

\bibitem{Maraspin04}
F.~Maraspin, P.~Bevilacqua, and P.~Rem, ``Modeling the throw of metals and
  nonmetals in eddy current separations,'' {\em International Journal of
  Mineral Processing}, vol.~73, no.~1, pp.~1--11, 2004.

\bibitem{Lungu01}
M.~Lungu and Z.~Scheltt, ``Vertical drum eddy-current separator with permanent
  magnets,'' {\em International Journal of Mineral Processing}, vol.~63,
  pp.~207--216, 2001.

\bibitem{Ruan11}
J.~Ruan and Z.~Xu, ``A new model of repulsive force in eddy current separation
  for recovering waste toner cartridges,'' {\em Journal of Hazardous
  Materials}, vol.~192, no.~1, pp.~307--313, 2011.

\bibitem{Coey10}
J.~M.~D. Coey, {\em Magnetism and Magnetic Materials}.
\newblock New York: Cambridge University Press, 2010.

\bibitem{OHandley00}
R.~C. O'{H}andley, {\em Modern Magnetic Materials: Principles and
  Applications}.
\newblock Hoboken, NJ: Wiley, 2000.

\bibitem{Arthur08}
J.~W. Arthur, ``The fudamentals of electromagnetic theory revisited,'' {\em
  IEEE Antennas and Propagation Magazine}, vol.~50, no.~1, pp.~19--65, 2008.

\bibitem{Powers99}
D.~L. Powers, {\em Boundary Value Problems}.
\newblock London: Academic Press, 4~ed., 1999.

\bibitem{Jackson99}
J.~D. Jackson, {\em Classical Electrodynamics}.
\newblock Hoboken, NJ: Wiley, 3~ed., 1999.

\bibitem{Ray17}
J.~D. Ray, J.~R. Nagel, D.~Cohrs, and R.~K. Rajamani, ``Forces on particles in
  time-varying magnetic fields,'' {\em {KONA} Powder and Particle Journal},
  vol.~(accepted for publication), 2017.

\bibitem{Nagel18}
J.~R. Nagel, ``Induced eddy currents in simple conductive geometries,'' {\em
  IEEE Antennas and Propagation Magazine}, vol.~(accepted for publication),
  pp.~0--10, 2018.

\end{thebibliography}

\end{document}